\documentclass[twocolumn,
amsmath,amssymb,
aps,prl,showpacs
]{revtex4-1}

\usepackage{graphicx}
\usepackage{bm}
\usepackage{epstopdf}
\usepackage{color}
\begin{document}

\title{3D spatially-resolved optical energy density enhanced by wavefront shaping}

\author{Peilong Hong}
\altaffiliation{Both authors contributed equally to this work.}
\email{Present address: School of Science, Nanjing University of Science and Technology, Nanjing, Jiangsu, China}
\author{Oluwafemi S. Ojambati}
\altaffiliation{Both authors contributed equally to this work.}
\author{Ad Lagendijk}
\author{Allard P. Mosk}
\altaffiliation{Present address: Nanophotonics, Debye Institute, Utrecht University, P.O. Box 80.000, 3508 TA Utrecht, The Netherlands}
\author{Willem L. Vos } 
\email{Corresponding author: w.l.vos@utwente.nl}

\affiliation{Complex Photonic Systems (COPS), 
MESA+ Institute for Nanotechnology, 
University of Twente, 
P.O. Box 217, 7500 AE Enschede, The Netherlands}

\date{23 March 2017} 

\begin{abstract}
We study the three-dimensional (3D) spatially-resolved distribution of the energy density of light in a 3D scattering medium upon the excitation of open transmission channels. 
The open transmission channels are excited by spatially shaping the incident optical wavefronts. 
To probe the local energy density, we excite isolated fluorescent nanospheres distributed inside the medium. 
From the spatial fluorescent intensity pattern we obtain the position of each nanosphere, while the total fluorescent intensity gauges the energy density. 
Our 3D spatially-resolved measurements reveal that the local energy density versus depth $(z)$ is enhanced up to $26 \times$ at the back surface of the medium, while it strongly depends on the transverse $(x,y)$ position. 
We successfully interpret our results with a newly developed 3D model that considers the time-reversed diffusion starting from a point source at the back surface. 
Our results are relevant for white LEDs, random lasers, solar cells, and biomedical optics. 

\end{abstract}

\maketitle


The interference of multiple scattered waves in complex media holds much fascinating physics such as coherent backscattering, Anderson localization, and mesoscopic correlations~\cite{vanRossum1999RMP, Sheng2006Book, Akkermans2007Book, Wiersma2013NP}. 
Transport through complex media is described by so-called channels that are eigenmodes of the transmission matrix~\cite{Beenakker1997RMP}. 
Remarkably, open transmission channels with near-unity transmission are predicted to perfectly transmit a properly designed incident field even if the medium is optically thick~\cite{Dorokhov1984SSC}. 
It has recently been demonstrated that light is sent into open transmission channels by a spatial shaping of the incident wavefronts~\cite{Vellekoop2008PRL,Popoff2014PRL,Mosk2012NP}. 
This development has led to tightly focused transmitted light (henceforth referred to as ``optimized light'')~\cite{Vellekoop2007OL,Vellekoop2010NP,Katz2011NP,vanPutten2011PRL}, enhanced optical transport through a scattering medium~\cite{Vellekoop2008PRL,Kim2012NP,Popoff2014PRL,ChoiSciRep2015,Ojambati2016PRA}, and imaging through~\cite{Popoff2010NC,Katz2012NP,Bertolotti2012N} and even inside a scattering medium~\cite{Horstmeyer2015NP}. 

In contrast, only a few studies address the energy density of optimized light that plays a central role in applications of light-matter interactions, such as solid-state lighting~\cite{phillips2007LaserPhotonRev}, random lasers~\cite{Lawandy1994Nature}, solar cells~\cite{Polman2012Nature}, and biomedical optics~\cite{Wang2012biomedical}. 
In absence of wavefront control, the ensemble-averaged energy density depends linearly on depth $z$ in the medium~\cite{vanRossum1999RMP}. 
A critical question is how the energy density can be controlled by exciting open channels, and what the resulting 3D energy density is. 
In particular, the 3D energy density profile of shaped light has not been experimentally studied to date.
Due to the inherent opacity, direct optical imaging cannot be used 
to probe the 3D energy density profile. 
In Ref.~\cite{Ojambati2016NJP}, it was shown that spontaneous emission of embedded fluorescent nanoparticles does report the energy density and it was observed that the depth-integrated \textit{global} energy density is increased by wavefront shaping but unfortunately, the 3D profile was unresolved. 
Several studies on low-dimensional systems~\cite{Choi2011PRB,Davy2015NC,Liew2015OE,Sarma2016PRL,Ojambati2016OE,Koirala2017arXiv} indicate that the energy density versus $z$ position has a maximum near the center of the sample, while the transverse $(x,y)$ behavior remained not addressed. 
Thus, to investigate how the 3D local optical energy density is controlled by wavefront shaping, a \textit{local} 3D $(x,y,z)$-resolved measurement is called for. 

\begin{figure}[htbp]
\center
\includegraphics[width=1.0\columnwidth]{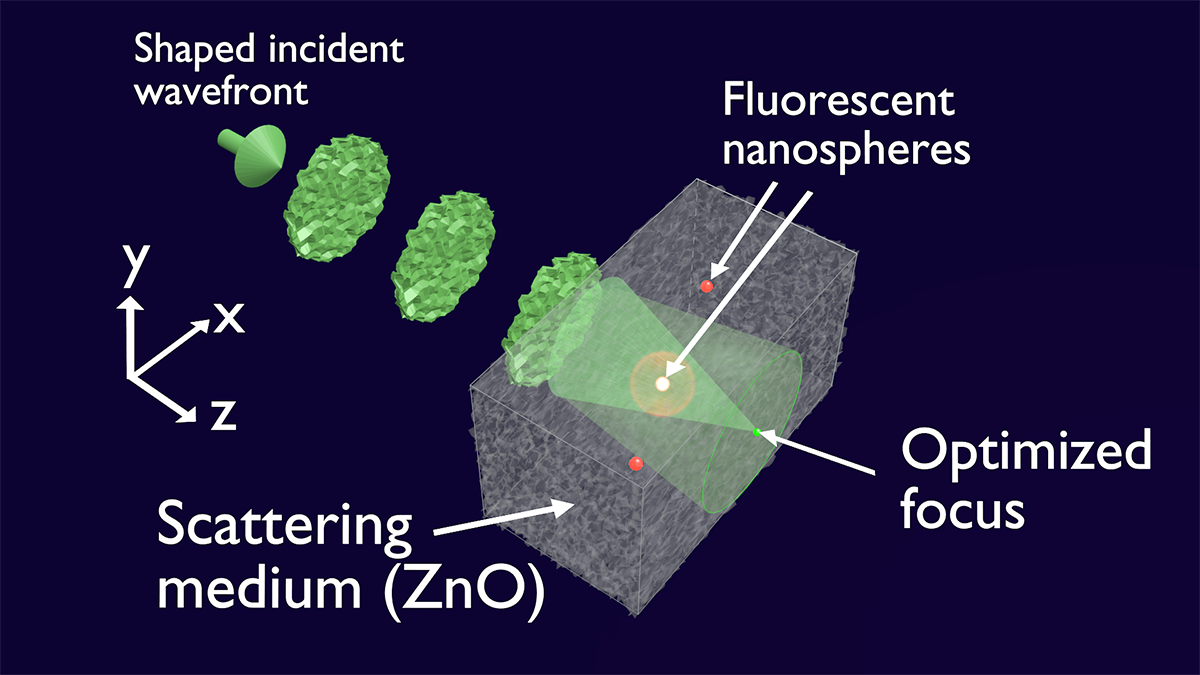}
\caption{(Color) 
Method to probe the 3D $(x,y,z)$ spatially-resolved local energy density that is enhanced by wavefront shaping. 
Incident green light is wavefront shaped to an optimized focus at the back surface of a scattering medium (ensemble of ZnO nanospheres) to excite open transmission channels. 
The scattering medium is sparsely doped with fluorescent nanospheres that probe the local energy density of the green incident light at different positions $(x,y,z)$ by emitting orange light in proportion to the local energy density of the green excitation light.  }
\label{fig:figure1}
\end{figure}

In this work, we investigate the 3D \textit{local} spatially-resolved energy density in a 3D scattering medium, with optimized incident light.
Figure~\ref{fig:figure1} illustrates our experiment: using a spatial light modulator (SLM), we shape the incident green light to a focus at the back surface of a disordered ensemble of ZnO nanoparticles, a procedure that is known to couple light into open channels~\cite{Vellekoop2008PRL,Choi2011PRB,Ojambati2016OE}.
The resulting energy density is probed \textit{locally} by fluorescent nanospheres. 
The density of the nanospheres is so low that only one of them is present in the illuminated volume.
Wavefront shaping increases the local energy density by an enhancement factor that we denote as $\eta_{\rm{f}}(x,y,z)$. 
Consequently, the fluorescence emission of a nanosphere, which is proportional to the local energy density at its location, is enhanced by the same factor.
We performed measurements on several nanospheres inside a sample and for each individual sphere we measured two key parameters, namely the nanosphere location $(x,y,z)$ and the differential fluorescence enhancement $\partial\eta_{\rm{f}}/\partial F$. 
Here  the fidelity $F$ quantifies the overlap between the experimentally-generated wavefront and the perfect wavefront that optimally couples light to the target position~\cite{Vellekoop2008PRL}.  

\begin{figure}[tbp]
\center
\includegraphics [width=1.0\columnwidth] {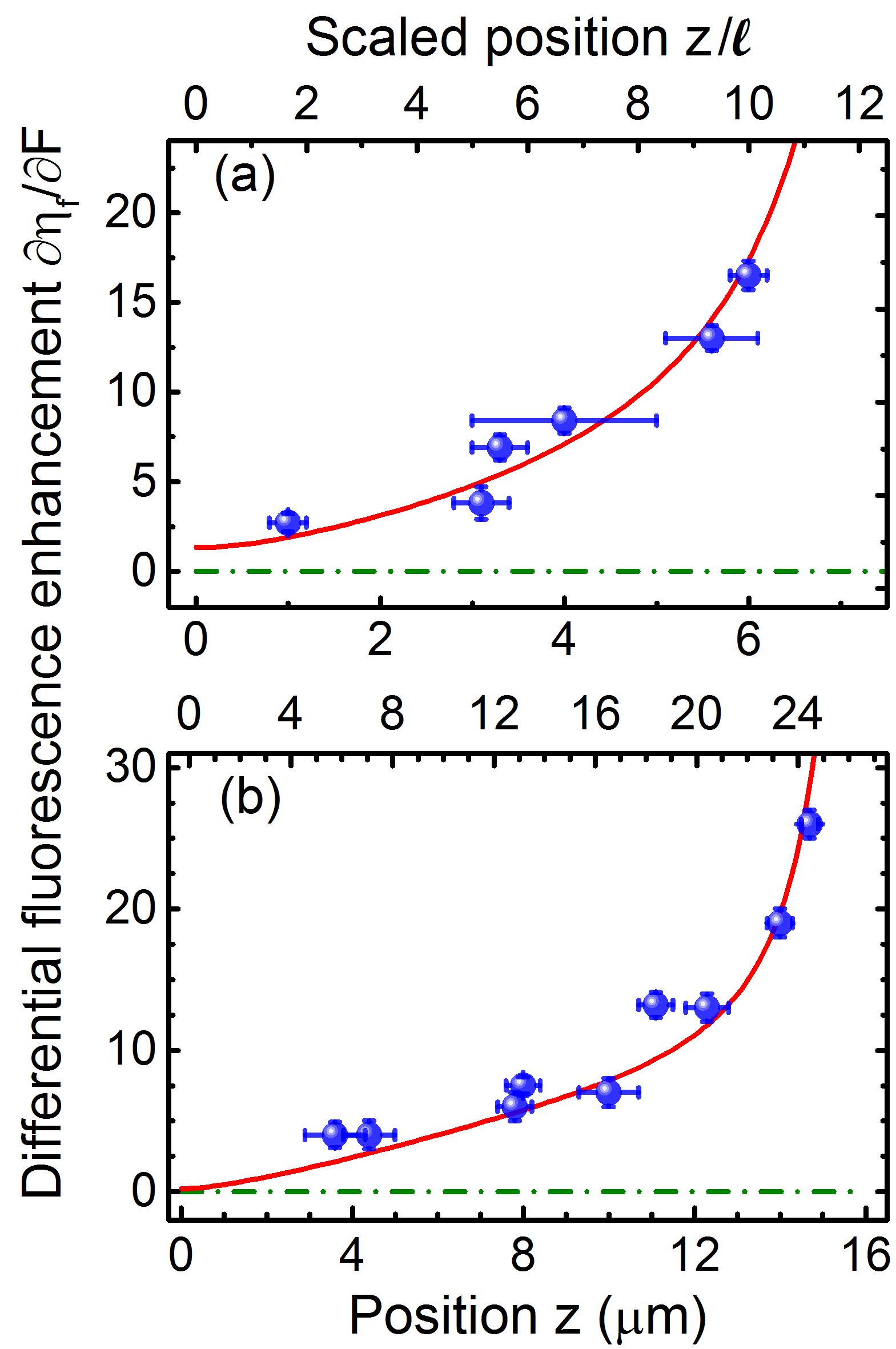}
\caption{(Color online.) 
Measured differential fluorescence enhancement $\partial\eta_{\rm{f}}/\partial F$ versus position $z$ and scaled position $z/\ell$ for two samples, each with thicknesses (a) $L = 8~\mu$m and (b) $L = 16~\mu$m. 
The fluorescent nanospheres are centered at the optical axis at $(x_0, y_0) = (0, 0)$.
Blue circles are the measured results with error bars. 
The red curve is the energy density enhancement predicted by our 3D model.
The green dash-dotted curve indicates zero energy density enhancement.   }
\label{fig:enhancement_vs_depth}
\end{figure}

\begin{figure}[tbp]
\center
\includegraphics [scale=0.45] {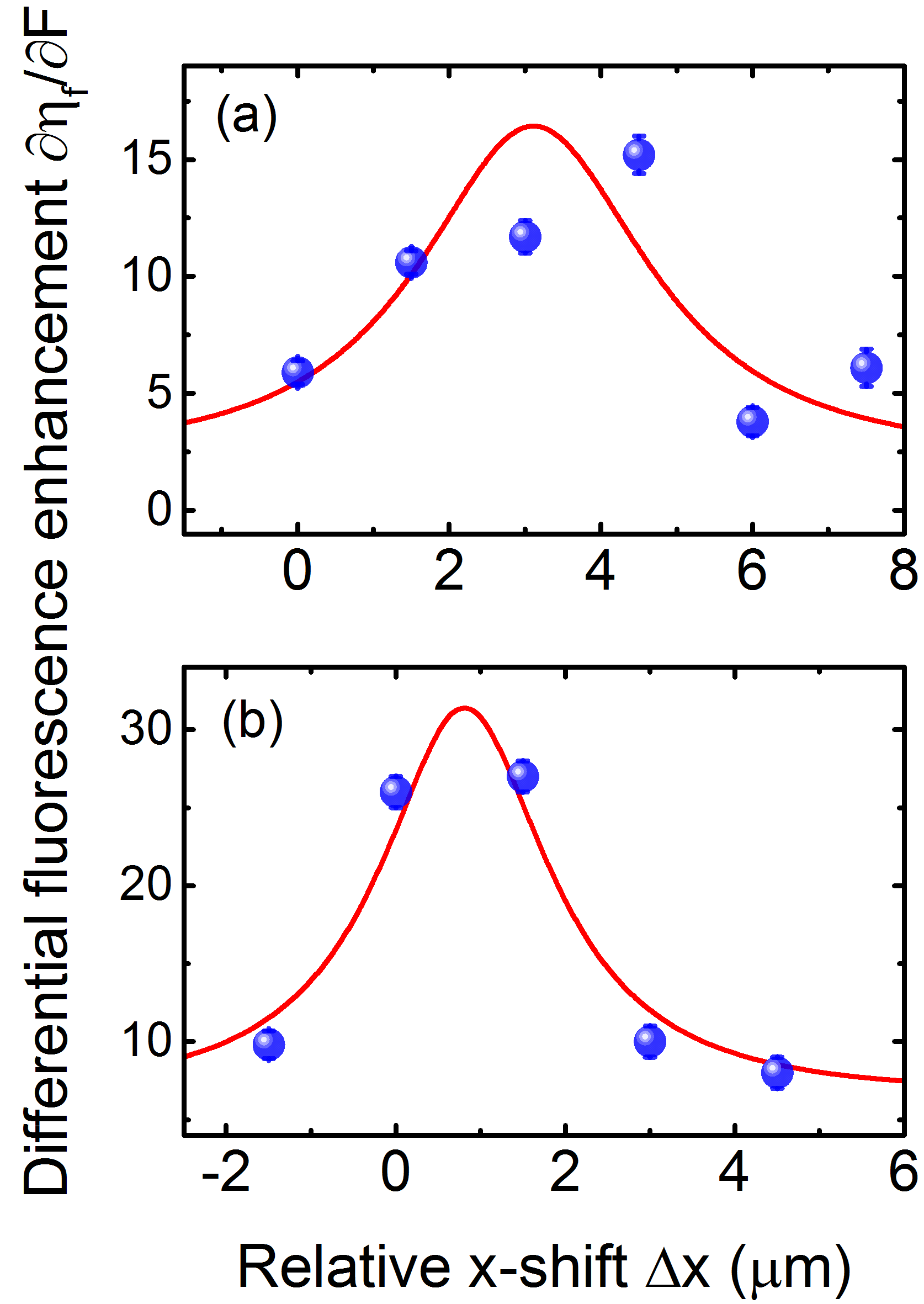}
\caption{(Color online.) 
Differential fluorescence enhancement $\partial \eta_{\rm{f}}/\partial F$ versus transverse displacement $\Delta x$ relative to the optical axis for scattering samples with thicknesses $L = 8 \, \mu$m (a) and $L = 16 \, \mu$m (b).  
Blue circles are the measured data with error bars. 
Red lines are the energy density enhancement predicted by our 3D model. 
In (a) and (b), the selected fluorescence nanospheres are located at $(y, z)$ = $(0,\, 5.9 \,\pm \,0.1 )\,\mu$m and $(y, z)$ = $(0, \, 14.7 \, \pm \, 0.2 )\, \mu$m, respectively.
}
 \label{fig:figure7}
\end{figure}

Fig.~\ref{fig:enhancement_vs_depth} and \ref{fig:figure7} show our main results: the measured differential fluorescence enhancement $\partial\eta_{\rm{f}}/ \partial F$ versus $z$ and $x$ positions, respectively, in scattering samples with thicknesses $L = 8 \, \pm \, 2 \,\mu$m and $16 \, \pm \, 2 \,\mu$m. 
In Fig.~\ref{fig:enhancement_vs_depth}, the differential fluorescence enhancement $ \partial \eta_{\rm{f}}/ \partial F$ increases with $z$ position from front to back, and $ \partial \eta_{\rm{f}}/ \partial F$ increases up to $16$ and $26$ with thickness.
The data deviates significantly from the uncontrolled limit $(\partial\eta_{\rm{f}}/ \partial F = 0)$, which reveals that wavefront shaping of light changes the local energy distribution.
We propose a 3D model without free parameters that describes the data in Fig.~\ref{fig:enhancement_vs_depth} very well. 

To verify the 3D character of $\eta_{\rm{f}}(x,y,z)$, we translate the sample along the $x$-axis while keeping $(y,z)$ constant.
In Fig.~\ref{fig:figure7}, we plot the differential fluorescence enhancement $\partial \eta_{\rm{f}}/\partial F$ versus the $x$-displacement $\Delta x$ relative to the optical axis $(x_0, y_0)$.
For both samples, $\partial \eta_{\rm{f}}/\partial F$ reveals clear maxima, revealing the effect of the optimized focus.
Due to $(x,y)$ symmetry in the transverse plane, a similar behavior occurs versus $y$ coordinate, and requires considerable acquisition time (see Supplementary~\cite{footnote:EPAPS}). 
In Figs~\ref{fig:figure7}(a) and (b), the maxima are centered at $\Delta x = 3\,\mu \rm{m}$ and $\Delta x = 1 \,\mu\rm{m}$ respectively, due to a slight displacement of the nanosphere from the optical axis at $x_0 = 0$. 
This observed strong dependence on $x$ is also well described by our 3D model and is not explained by the 1D diffusion model that is necessarily independent of $(x,y)$-coordinate. 

\begin{figure}[htbp]
\center
\includegraphics [width=1.0\columnwidth] {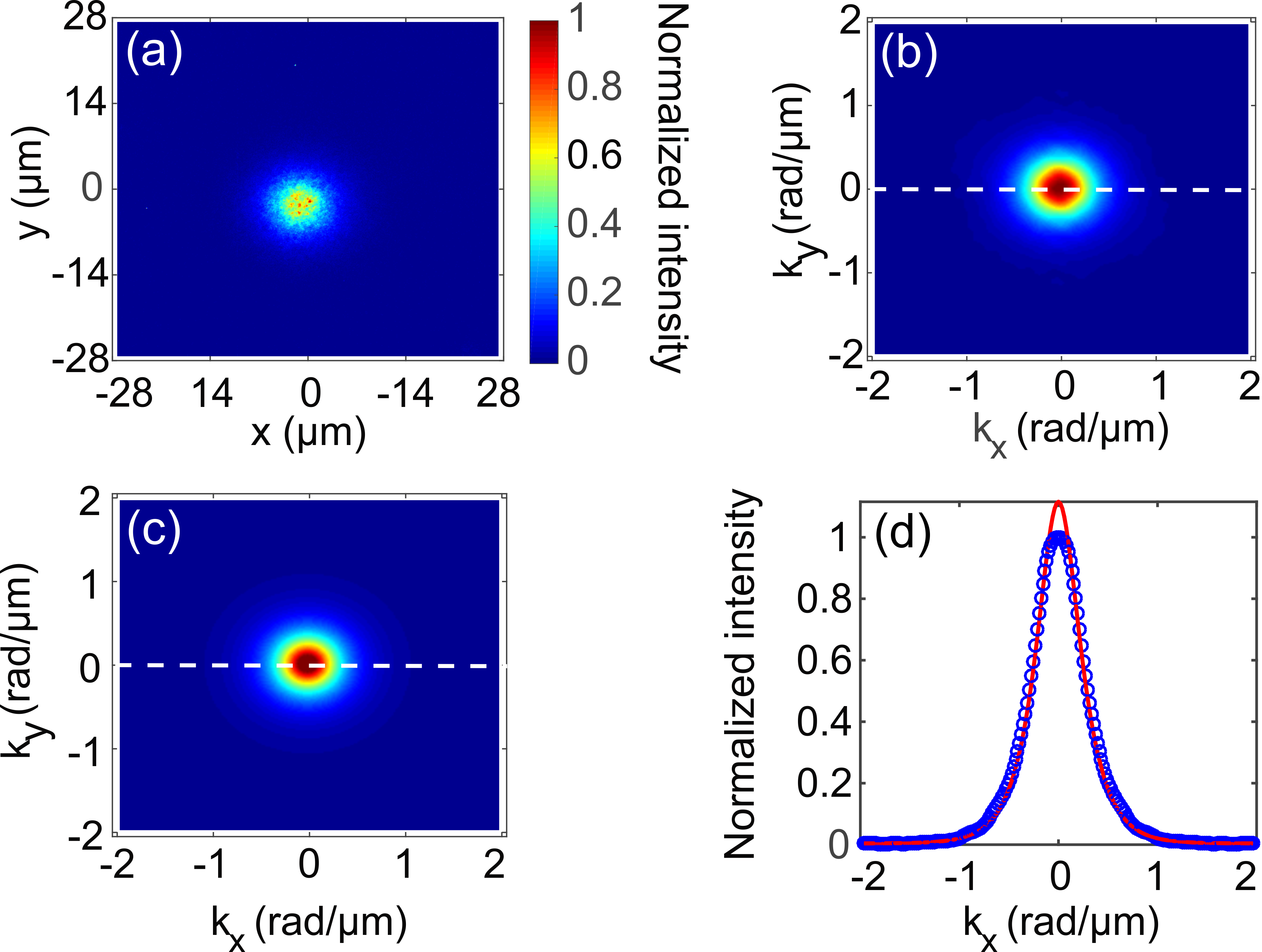}
\caption{(Color online) 
Determining the position $z$ of a fluorescent sphere in the $L = 8~\mu$m thick sample. 
(a) Fluorescence image measured in real space by averaging 41 data sets, each with a different random incident wavefront. 
(b) Fourier transform of (a). 
Intensities in (a) and (b) were normalized to their maxima. 
(c) Solution of the diffusion equation with a fluorescent nanosphere at position $z = 3.3~\pm 0.3~\mu$m.
(d) Cross sections through the centers of (b) (blue circles) and (c) (red line), respectively. 
}
\label{fig:figure_depth_fitting}
\end{figure}

The samples were prepared by spray-painting a suspension of ZnO nanoparticles and a low concentration of fluorescent particles on a glass slide (see Supplementary for details). 
A dense ensemble of strongly scattering ZnO nanoparticles was obtained after evaporation.
The locations $(x,y,z)$ of the fluorescent nanospheres are \textit{a priori} unknown since the nanospheres end up at random positions. 
To determine the locations $(x,y,z)$, we first scanned the sample to find isolated fluorescent spheres, and then recorded the diffuse fluorescent spot at the back surface of the sample (see Fig.~\ref{fig:figure_depth_fitting}(a)). 
We performed a Fourier transformation of the fluorescent spot (see Fig.~\ref{fig:figure_depth_fitting}(b)) and filtered high-frequency noise. 
We model the nanosphere as a point source in the 3D diffusion equation~\cite{Vellekoop2008OE} and fit the solution in Fourier space to the Fourier transform of the fluorescence spot with  the nanosphere position $z$ as the only adjustable parameter, see Fig.~\ref{fig:figure_depth_fitting}(c-d). 

Next, we performed wavefront shaping experiments with the optical axis of the system centered on a nanosphere at coordinates $(x_0,y_0)$.
We obtained a feedback signal for the wavefront shaping optimization from an area of $0.03\,\mu m^2$ less than the speckle area $A = \lambda_{\rm{e}}^2/(2 \pi) = 0.05\,\mu m^2$.
We used the piecewise sequential algorithm to find the optimized incident wavefront~\cite{Vellekoop2007OL,Vellekoop2008PRL}, with 900 input degrees of freedom on the spatial light modulator.

Ideally, a perfectly shaped wavefront is the phase conjugate of the wavefront originating from a point source located at the target position~\cite{Vellekoop2008PRL}. 
A real wavefront in an experiment inevitably differs from a perfect wavefront due to finite resolution, temporal decoherence, modulation noise, and spatial extent of the generated field~\cite{Vellekoop2007OL,Vellekoop2008PRL,yilmaz2013BOE}.
The deviation of the wavefront from the ideal one due to all these effects can be represented in a single measure, the fidelity $F$.
Experimentally, the fidelity $F$ is gauged as $F = {n^2 \cdot I_{\rm{opt}}}/{I_{\rm{tot}}}$, where $n$ is the refractive index of the substrate, $I_{\rm{opt}}$ the intensity for the optimized wavefront, and $I_{\rm{tot}}$ the total transmitted intensity with an unoptimized reference wavefront~\cite{Vellekoop2008PRL,Ojambati2016NJP}. 
Since a real wavefront is the superposition of the perfectly shaped wavefront that controls the energy density and a random error wavefront~\cite{Vellekoop2008PRL}, the energy density $W_{\rm{e}} (x,y,z)$ due to a real incident wavefront is necessarily a linear combination of the perfectly optimized energy density $W_{\rm{o}}(x,y,z)$ and a diffusive unoptimized energy density $W_{\rm{uo}}(x,y,z)$
\begin{equation}\label{eq:W_real}
W_{\rm{e}}(x,y,z) = F . W_{\rm{o}}(x,y,z) + (1 - F) .  W_{\rm{uo}}(x,y,z).
\end{equation}
(The energy densities in Eq.~\ref{eq:W_real} are ensemble averaged over several realizations). 
By probing the fluorescent spheres at different positions, we obtain the local energy density enhancement defined as $\eta_{\rm{f}}(x,y,z) \equiv W_{\rm{e}}(x,y,z) / W_{\rm{uo}}(x,y,z)$.
Eq.~(\ref{eq:W_real}) leads to a linear dependence of the energy density enhancement on fidelity
\begin{equation}\label{eq:EnhFid}
\eta_{\rm{f}}(x,y,z) = 1 + F . \frac{\partial \eta_{\rm{f}}}{\partial F} 
\end{equation}
with \emph{unity} intercept and a slope $\partial \eta_{\rm{f}}/\partial F = (W_{\rm{o}}/W_{\rm{uo}} - 1)$ that we call the \textit{differential fluorescent enhancement}.

\begin{figure}[htbp]
\center
\includegraphics[width=0.9\columnwidth]{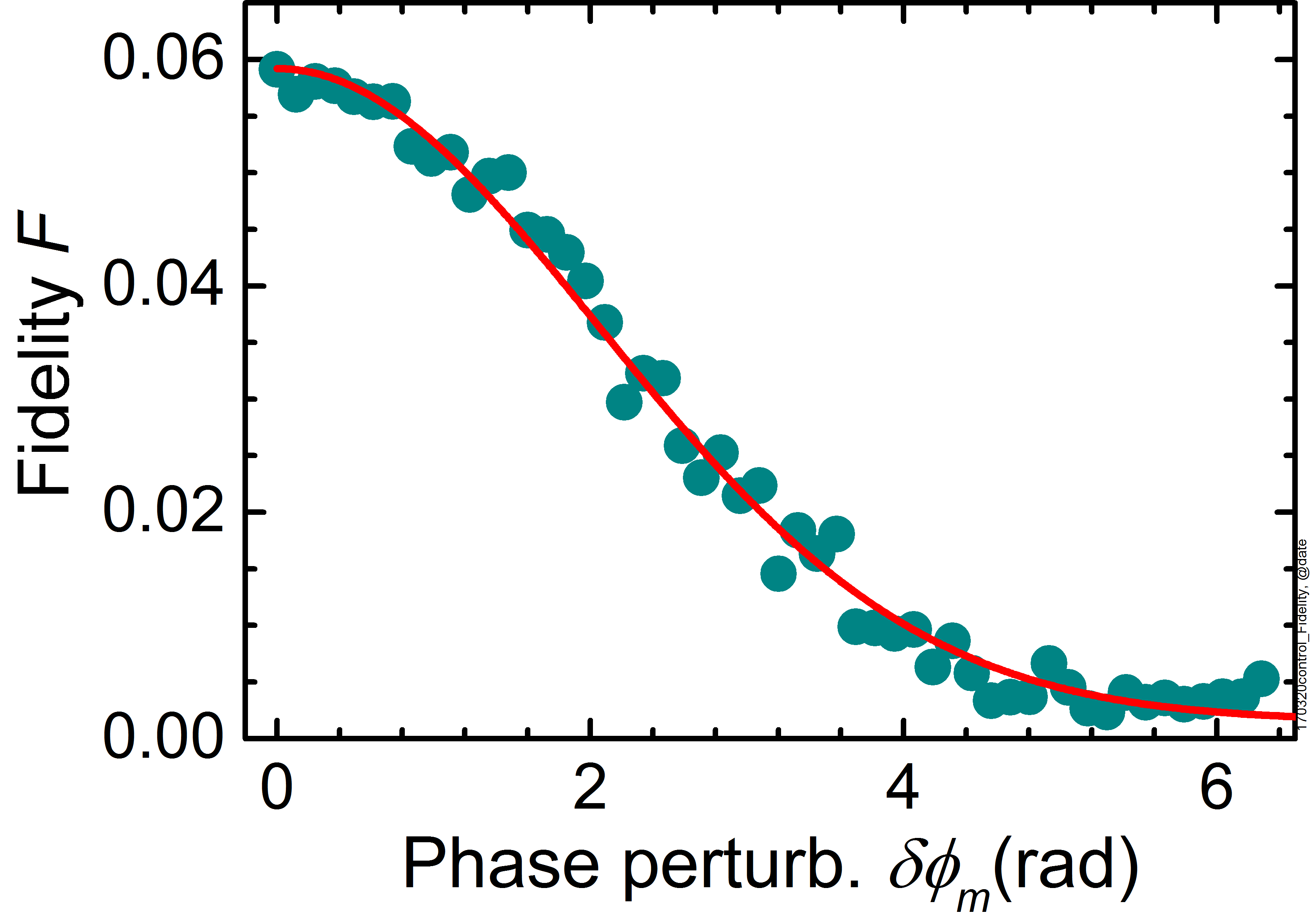}
\caption{(Color online). 
Measured fidelity $F$ versus the phase perturbation factor $\delta \phi_{\rm{m}}$ on the optimized phase pattern. 
We applied $m_1 = 41$ phase perturbations to each optimal wavefront. 
The red curve is a guide to the eye. }
\label{fig:figure3}
\end{figure}

To determine $\partial \eta_{\rm{f}}/\partial F$ from Eq.~\ref{eq:EnhFid}, we scanned the fidelity $F$ by intentionally adding a random perturbation phase to each segment of the optimized wavefront. 
Figure~\ref{fig:figure3} shows that the fidelity $F$ continuously decreases with increasing phase perturbation $\delta \phi_{\rm{m}}$ from the maximum obtained fidelity to $0$. 
For each perturbed phase, we collected fluorescence images $I_{\rm{p}}$.
We also collected 41 reference fluorescent images $I_{\rm{r}}$ each with a random phase pattern as the input wavefront. 
We determined experimentally the fluorescence enhancement $\eta_{\rm{f}}$ from the ratio of $ I_{\rm{p}}$ and the average $I_{\rm{r}}$.
We repeated the wavefront shaping and fidelity scanning procedure $m_2 = 100$ times on each nanosphere to obtain an ensemble average.  

\begin{figure}[htbp]
\center
\includegraphics[width=0.9\columnwidth]{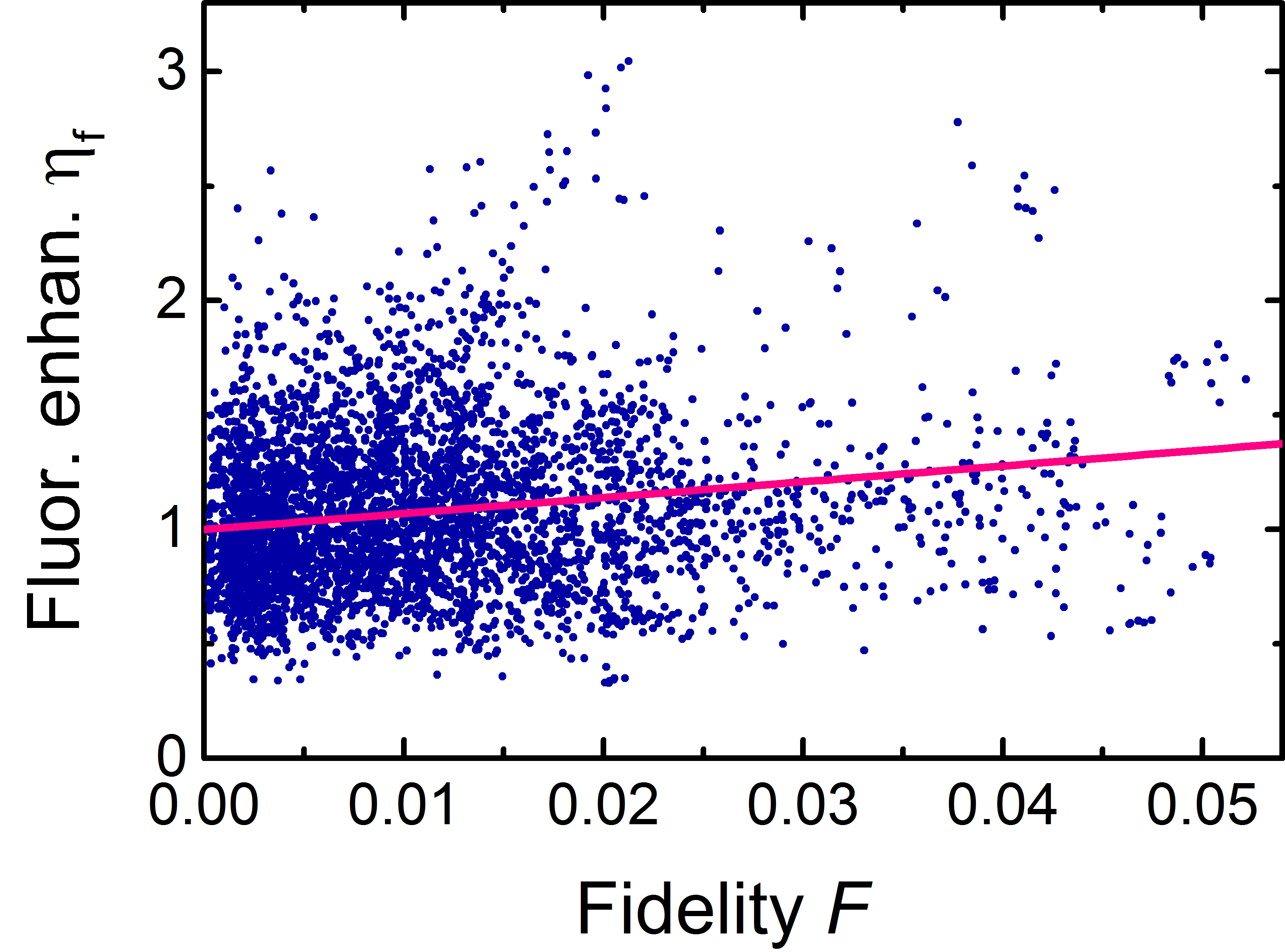}
\caption{(Color online.) 
Measured fluorescence enhancement $\eta_{\rm{f}}$ versus fidelity $F$ (blue dots) at a single position $z = 3.3\pm 0.3 \mu$m in the $L = 8~\mu$m thick sample. 
The red line from Eq.~\ref{eq:EnhFid} with unity intercept and the only adjustable parameter the differential fluorescence enhancement $\partial \eta_{\rm{f}}/\partial F$ that is determined to be $ 6.9 \pm 0.7$, where the error is the 95\% confidence interval. 
The confidence interval is within the line thickness. 
}
\label{fig:figure4}
\end{figure}

The measured collection of $(m_1.m_2 = 4100)$ fluorescence enhancement $\eta_{\rm{f}}$ data points versus fidelity $F$ is shown in Fig.~\ref{fig:figure4} for one fluorescent nanosphere. 
While the data show inevitable variations (see Supplementary~\cite{footnote:EPAPS}), the fluorescence enhancement clearly increases with $F$, to an average of $1.4 \times$ at the maximum obtained fidelity. 
From the linear dependence between $\eta_{\rm{f}}$ and $F$ with unity axis intercept described by Eq.~\ref{eq:EnhFid}, we obtain the slope $\partial \eta_{\rm{f}}/\partial F$ at a specific position $z$ on the optical axis $(x_0,y_0)$. 
All obtained $\partial \eta_{\rm{f}}/\partial F$ for the two samples are shown in Figure~\ref{fig:enhancement_vs_depth}. 
The procedure described above was repeated at various $\Delta x$ displacements and the results are shown in Fig~\ref{fig:figure7}.

To model the 3D energy density $W_{\rm{o}}(x,y,z)$ of optimized light, we consider the optimized target to be a point source of diffuse light, as shown in Fig.~\ref{fig:figure1}. 
The 3D energy density $W_{\rm{dif}}(x,y,z)$ of the point source is described by the 3D diffusion equation~\cite{vanRossum1999RMP,Vellekoop2008OE} (see Supplementary for details~\cite{footnote:EPAPS}).
Light from the point source diffuses in a cone from the back surface to the front surface via open channels. 
While the time reverse (or phase conjugate) of the light transmitted to the front surface describes light traveling to the target point at the back surface, part of the light injected at the front surface contributes to a background, notably in the space outside the optimized focus (see Fig.~\ref{fig:figure1}). 
The background light is caused light coupled to closed channels that are mainly reflected and thus corresponds to an incomplete time reversal (or phase conjugation) of light from the optimized target.
At perfect fidelity, we thus describe the optimized energy density $W_{\rm{o}}(x,y,z)$ as a sum of two components $W_{\rm{o}}(x,y,z) = W_{\rm{of}}(x,y,z) + W_{\rm{bg}}(x,y,z)$, with $W_{\rm{of}}$ the energy density originating from the optimized focus, and $W_{\rm{bg}}$ the background energy density. 
Following the maximal fluctuation approximation by Pendry \textit{et al.}, we describe $W_{\rm{of}}$ and $W_{\rm{bg}}$ by assuming the transmission channels to consist of only open and closed channels~\cite{Pendry1990PhysA,Pendry1992MathProc}.
For open channels, it has recently been shown that the energy density profile along $z$ tracks the fundamental mode of the diffusion equation 
~\cite{Ojambati2016NJP,Ojambati2016OE,Koirala2017arXiv}.
To obtain $W_{\rm{of}}$, we therefore normalize $W_{\rm{dif}}(x,y,z)$ and map it onto the spatial profile of the fundamental diffusion mode (see Supplementary~\cite{footnote:EPAPS}). 
Similarly, we describe $W_{\rm{bg}}$ by mapping the fundamental diffusion mode onto a Gaussian profile with a constant width along $z$. 
The amplitudes of $W_{\rm{fo}}$ and $W_{\rm{bg}}$ are fixed by the total transmitted intensity.

In our experiments, a fluorescent nanosphere at position $(x,y,z)$ is excited by the local energy density $W_{\rm{o}}(x, y, z)$ in case of optimized light (modeled above), or $W_{\rm{uo}}(x,y,z)$ for unoptimized incident light. 
We describe $W_{\rm{uo}}(x,y,z)$ as a product of the solution of the 1D diffusion equation (versus $z$) and a Gaussian (in $(x,y))$. 
From the ratio of $W_{\rm{o}}$ and $W_{\rm{uo}}$, we obtain $\partial \eta_{\rm{f}}/\partial F$ that we plot as a function of $z$ in Fig.~\ref{fig:enhancement_vs_depth}.
For both samples, our 3D model shows that $\partial \eta_{\rm{f}}/\partial F$ increases steadily as $z$ increases to the back surface of the sample, 
in excellent agreement with the experimental data. 
This agreement shows that the intensity enhancement observed on the back surface is associated with the 3D enhancement of the local energy density in the bulk. 

In summary, by exciting open transmission channels in a 3D scattering medium by wavefront shaping, we observe that the local energy density is considerably enhanced. 
The enhancement increases towards the back surface of the sample and has a maximum along the transverse direction, revealing the effect of the optimized focus. 
A 3D model without adjustable parameters successfully describes the experimental data.
Our results thus offer new insights on the 3D redistribution of the energy density in 3D scattering media, which is extremely useful to enhance the efficiency of energy conversion in systems such as random lasers, solar cells, and white LEDs.
For white LEDs, wavefront shaping could serve to control the color temperature by optimizing for ``warm'' or ``cold'' white light. 
Our results are also applicable to wavefront shaping of classical waves such as acoustic and pressure waves~\cite{Derode1995PRL}, and to quantum waves such as electrons in nanostructures.

\begin{acknowledgments}
We thank Cornelis Harteveld  for technical assistance, and Jeremy Baumberg, Sylvain Gigan, Pepijn Pinkse, Stefan Rotter, Tristan Tentrup, Wilbert IJzerman (Philips Lighting) and Floris Zwanenburg for helpful discussions. 
This work is part of the research program of the ``Stichting voor Fundamenteel Onderzoek der Materie'' (FOM) FOM-program ``Stirring of light!'', which is part of the ``Nederlandse Organisatie voor Wetenschappelijk Onderzoek'' (NWO), and we acknowledge support by NWO-Vici, DARPA, and STW.
\end{acknowledgments}

\end{document}